\documentclass[11pt,twoside]{article}
\usepackage{cspm2015}

\usepackage{hyperref}

\aspSuppressVolSlug
\resetcounters

\def\figspath{.} 

\newcommand{\jgr}{JGR}
\newcommand{\asr}{{Adv. Space Res.}}
\newcommand{\pop}{{Phys. Plasmas}}

\begin{document}

\title{Modeling Jets in the Corona and Solar Wind}
\author{T.~T\"or\"ok,$^1$ R.~Lionello,$^1$ V.~S.~Titov,$^1$ J.~E.~Leake,$^2$ Z.~Miki\'c,$^1$ J.~A.~Linker,$^1$ and M.~G.~Linton$^2$
\affil{$^1$Predictive Science Inc., 9990 Mesa Rim Rd., Ste. 170, San Diego, CA 92121, USA; \email{tibor@predsci.com}}
\affil{$^2$US Naval Research Laboratory 4555 Overlook Ave., SW Washington, DC 20375, USA}}

\paperauthor{T.~T\"or\"ok}{tibor@predsci.com}{ORCID_Or_Blank}{Author1 Institution}{Author1 Department}{City}{State/Province}{Postal Code}{Country}

\begin{abstract}
Coronal jets are transient, collimated eruptions that occur in regions of predominantly open magnetic field in the solar corona. Our understanding of these events has greatly evolved in recent years but several open questions, such as the contribution of coronal jets to the solar wind, remain. Here we present an overview of the observations and numerical modeling of coronal jets, followed by a brief description of ``next-generation'' simulations that include an advanced description of the energy transfer in the corona (``thermodynamic MHD''), large spherical computational domains, and the solar wind. These new models will allow us to address some of the open questions.
\end{abstract}

%
\section{Introduction}
%
Coronal (or X-ray) jets are transient, collimated eruptions of hot plasma which may significantly contribute to heating the corona and powering the solar wind. They occur in regions of open or semi-open magnetic fields in the solar corona and are believed to be launched by magnetic reconnection between closed and open fields, as a result of flux emergence or flux cancellation. While reconnection appears to be essential for the formation of jets, the detailed physical mechanisms that drive plasma heating and ejection in these events are still under debate \citep[e.g.,][]{cheung15}. Similar phenomena (e.g., chromospheric jets and Type-II spicules) occur on smaller scales and in larger numbers in the low atmosphere and may be driven by analogous mechanisms \citep{takasao13} or solely by Alfv\'en waves \citep{cranmer15}. In this article we restrict our attention to the observations and the numerical modeling of coronal jets.     

\begin{table}[t]
\[
\resizebox{\textwidth}{!}{%
\begin{tabular}{|l|l|l|l|l|l|}
\hline
\multicolumn{1}{|c|}{Jet property} & \multicolumn{5}{c|}{Instrument} \\
& \multicolumn{1}{c}{SXT$^1$}  
& \multicolumn{1}{c}{XRT$^2$}  
& \multicolumn{1}{c}{EUVI$^3$}
& \multicolumn{1}{c}{COR1$^4$}
& \multicolumn{1}{c|}{AIA$^5$}\\
\hline
\hline
Source region   &\begin{tabular}{lc}  Active region   \end{tabular}
                &\begin{tabular}{lc}  Polar CH        \end{tabular} 
                &\begin{tabular}{lc}  Polar/Equat. CH \end{tabular} 
                &\begin{tabular}{lc}  -----           \end{tabular} 
                &\begin{tabular}{lc}  Polar/Equat. CH \end{tabular} \\ \hline

Occurrence      &\begin{tabular}{lc} $\sim$\,17/month \end{tabular}
                &\begin{tabular}{lc} $\sim$\,60/day   \end{tabular}  
                &\begin{tabular}{lc} ----             \end{tabular} 
                &\begin{tabular}{lc} $\sim$\,15/day   \end{tabular} 
                &\begin{tabular}{lc} ----             \end{tabular} \\ \hline
           
Duration [min]  &\begin{tabular}{lc} 2\,--\,600       \end{tabular}
                &\begin{tabular}{lc} 5\,--\,40 (10)   \end{tabular}  
                &\begin{tabular}{lc} 20\,--\,40       \end{tabular}
                &\begin{tabular}{lc} $<$ 20\,--\,120  \end{tabular} 
                &\begin{tabular}{lc}  21\,--\,46      \end{tabular} \\ \hline
                
Velocity [km/s] &\begin{tabular}{lc} 10\,--\,1000 (200)\end{tabular}
                &\begin{tabular}{lc} 80\,--\,500  (160)\end{tabular}  
                &\begin{tabular}{lc} 270\,--\,400      \end{tabular}
                &\begin{tabular}{lc} 100\,--\,560 (270)\end{tabular} 
                &\begin{tabular}{lc} 94\,--\,760       \end{tabular} \\ \hline
                
Length [Mm]     &\begin{tabular}{lc} 30\,--\,400 (150) \end{tabular}
                &\begin{tabular}{lc} 10\,--\,120 (18)  \end{tabular}  
                &\begin{tabular}{lc} 100 (one event)   \end{tabular}
                &\begin{tabular}{lc} ----              \end{tabular} 
                &\begin{tabular}{lc} 63\,--\,188       \end{tabular} \\ \hline
            
Width [Mm]      &\begin{tabular}{lc} 5\,--\,100 (17)   \end{tabular}
                &\begin{tabular}{lc} 3\,--\,12 (7)     \end{tabular} 
                &\begin{tabular}{lc} 25 (one event)    \end{tabular}
                &\begin{tabular}{lc} ----              \end{tabular} 
                &\begin{tabular}{lc} ----              \end{tabular} \\ \hline                         
\end{tabular}}
\]
\caption{\small
Coronal jet properties as observed by different instruments. The numbers in parentheses 
are either average values or distribution peaks. CH stands for ``coronal hole''. Sources:
\citet{shimojo96}$^1$,
\cite{savcheva07}$^2$,
\citet{nistico09,nistico10}$^3$,
\citet{paraschiv10}$^4$,
\citet{moschou13}$^5$.
}
\end{table}

\subsection{Observations}
Coronal jets have been observed for several decades \citep[e.g.,][]{brueckner83}, but systematic studies started only after the launch of {\em Yohkoh}, when detailed soft X-ray observations from SXT became available \citep[e.g.,][]{shibata92,shimojo96, shimojo98}. Later studies also employed XRT data from {\em Hinode} \citep[e.g.,][]{savcheva07}, EUV data from {\em TRACE} \citep[e.g.,][]{alexander99}, {\em SOHO}/EIT \citep[e.g.,][]{ko05}, and now {\em SDO}/AIA \cite[e.g.,][]{moschou13}, white-light observations from the {\em SOHO}/LASCO and {\em STEREO}/COR coronagraphs \citep[e.g.,][]{wang.ym98,nistico10}, as well as spectroscopic data from {\em SOHO}/SUMER and {\em Hinode}/EIS \citep[e.g.,][]{culhane07,madjarska12}.

Such studies revealed that coronal jets typically exhibit an inverted-Y shape, consisting of an elongated spire and a set of small flaring loops (a ``bright point'') at their base (see Figure\,\ref{f:fig1} below). Larger events can last for tens of minutes, extend several hundred Mm or more into the corona, and reach plasma velocities of several hundred km\,s$^{-1}$. Table~1 shows a compilation of some coronal jet properties, obtained in different wavelengths by various instruments. Estimated plasma temperatures in jet outflows fall roughly into the range (0.5--2.0)\,MK \citep[e.g.,][]{culhane07,madjarska11} and seem to cluster around 1.5\,MK \citep[e.g.,][]{doschek10,nistico11,young14}. Blowout jets (see below) appear to be slightly hotter than standard jets \citep{pucci13}. Typical electron number densities are between $10^8\,\mathrm{and}\,10^9\,\mathrm{cm}^{-3}$ \citep[e.g.,][]{chandrashekhar14,paraschiv15}, but significantly higher values of about $10^{11}\,\mathrm{cm}^{-3}$ were reported for an active-region jet \citep{chifor08}. 

The observations have shown that coronal jets occur predominantly in regions of (fully or locally) open magnetic field, i.e., in polar and equatorial coronal holes and at the periphery of active regions. \citet{nistico10} conducted the first systematic study of jets in equatorial coronal holes and found strong indications that they have the same properties as polar jets. It was shown that coronal jets are typically associated with evolving mixed polarity regions in the photosphere, in particular with flux emergence \citep[e.g.,][]{shibata01} and flux cancellation \citep[e.g.,][]{chifor08}. Which of these two mechanisms is dominant in producing jets may strongly depend on the location where the jet takes place \citep{sako13}. Furthermore, it was found that coronal jets are closely related to other coronal and interplanetary phenomena, such as H$\alpha$ surges \citep[e.g.,][]{canfield96}, ``anemone active regions'' \citep[e.g.,][]{asai08}, plumes \citep{raouafi08,pucci14}, ``Mini-CMEs'' \citep[e.g][]{nistico09}, type III radio bursts, and SEP events \citep[e.g.,][]{nitta08}.

With the help of {\em STEREO}/EUVI, it was found that coronal jets often display untwisting motions, indicating a helical magnetic field structure and the occurrence of torsional Alfv\'en waves in many events \citep{nistico09}. Such ``twisting jets'' were seen earlier by {\em Yohkoh} \citep{shibata92}, but could now be studied in much more detail \citep{patsourakos08}. Coronal jets have been observed to be ``sympathetic'' \citep{pucci12}, as well as homologous or ``recurrent'', i.e., to occur several times at the same location, typically at the periphery of active regions \citep[e.g.,][]{guo.y13,cheung15}. Moreover, some observations indicate that coronal jets are not generated by a single impulsive energy release, but rather by a sequence of distinct reconnection events  \citep{madjarska11,pucci13,chandrashekhar14a}. 

In recent years a second type of coronal jet, so-called ``blowout'' jets, have attracted considerable attention. Compared to the ``standard'' jets described above such events exhibit a more elongated, arcade-like set of hot loops at their base, a spire that is wider and more complex, and are associated with the ejection of cool plasma ($10^4-10^5$\,K), which appears to be rarely the case for standard jets \citep{moore13}. \cite{moore10} suggested that blowout jets occur when a flux rope emerging into a unipolar coronal magnetic field carries sufficient twist to erupt, so that a standard jet (due to reconnection) may be followed by a blowout jet (due to eruption and reconnection). Such transitions are now regularly observed \citep[e.g.,][]{liu.c11,moore13}. 

Blowout jets can occur also without flux emergence, as a result of a ``regular'' eruption in a pre-existing fan-spine magnetic configuration \citep[e.g.,][]{su09,hong11}. They are often associated with (narrow) coronal mass ejections (CMEs). Sometimes the jet may just act as a trigger for an adjacent CME \citep[e.g.,][]{shen12}, while in other cases it evolves directly into a CME \citep[e.g.,][]{liu.j15,li.x15,thalmann15}. Since such events typically occur at the edges of active regions, they do not contradict the flux-emergence scenario for jets in coronal holes. However, \cite{sterling15} recently suggested, based on a sample of 20 polar events, that the reconnection that launches a jet may not be triggered by flux emergence, but by ``mini-filament'' eruptions originating at the inversion line between the parasitic polarity and the background field. Depending on the strength of the eruption, a standard or a blowout jet is produced. Sterling et al. speculated that continuous reconnection driven by flux emergence may not be sufficient to produce a jet, and that a dynamic ``burst'' of reconnection driven by an eruption is required \citep[see also][]{pariat09}.

A key discovery by {\em Hinode}/XRT was that coronal jets occur much more frequently than previously thought \citep[e.g.,][]{savcheva07}, suggesting that they may contribute to heating the corona and powering the solar wind \citep[e.g.,][]{cirtain07}. Jets have been observed to extend high into the corona, and \cite{neugebauer12} suggested that polar jets ultimately manifest as ``micro-streams'' observed in the fast solar wind. \cite{yu14} estimated the energy contribution by coronal jets to be only 1.6\,\% of the content required to heat the corona. \cite{paraschiv15} found an even smaller value but pointed out that many tenuous high-temperature jets may be missed by the observations. Indeed, \cite{young15} report the existence of ``dark jets'' that are seen in spectroscopic observations, but escape detection in EUV or soft X-ray. However, since it appears that dark jets are about as numerous as visible jets, their joint contribution to coronal heating may still remain small compared to the contribution that may be provided by low-atmospheric events \citep[e.g.,][]{depontieu11,moore11}.

\subsection{Numerical Modeling}
\label{ss:modeling}
Early on, after the first {\em Yohkoh} observations, it was suggested that coronal jets are triggered by reconnection between closed and open magnetic fields \citep{shibata92,shibata94}. The jet morphology (inverted Y-shape, often with a void between the bright point and the spire) indicates that the reconnection occurs around a magnetic null point. 

Based on this scenario, \citet{yokoyama95,yokoyama96} performed the first magnetohydrodynamic (MHD) simulation of a coronal jet, using a two-dimensional (2D) model of a buoyant, horizontal magnetic layer emerging into the corona in Cartesian geometry. They found that, indeed, a current sheet forms and subsequent reconnection occurs in a current sheet around a ``magnetic neutral point'' as the two flux systems press against each other. They pointed out that a significant inclination of the coronal field is necessary to produce an upward-directed jet. In their model, the jet is driven by thermal pressure gradients that form at the sites where the reconnection outflow collides with the pre-existing coronal field. Similar results were found later in extended 2D simulations of the model \citep[e.g.,][]{miyagoshi03,isobe07,nishizuka08}.

The first 3D MHD simulation of a coronal jet was presented more than a decade later by \citet{moreno-insertis08}. They used a Cartesian domain of the size of a small active region (about 30 Mm$^3$) and emerged a flux rope into a stratified atmosphere in hydrostatic equilibrium containing a uniform, tilted magnetic field. Their results are analogous to those of Yokoyama \& Shibata: current sheet formation, reconnection, and a hot jet produced by upward deflection of the reconnection outflow. This similarity is not trivial since in 3D different flux systems are, in principle, free to move around each other, thus avoiding reconnection. The main new ingredient in the simulation of  \citet{moreno-insertis08} is the 3D magnetic topology, which consists of a dome-like fan-separatrix surface a spine field line, both originating in a magnetic null point \citep[e.g.,][]{antiochos98,torok09}. The modeled jet properties are in very good agreement with the observations, except for the plasma temperature that seems too high (up to about  $10^7$\,K in the jet spire).  A similar simulation was performed by \citet{gontikakis09}, inspired by {\em TRACE} observations of an active-region jet. The main difference from the simulation by \cite{moreno-insertis08} is that the flux rope was emerged close to a small active region which was produced by a preceding flux-rope emergence. This simulation demonstrated how coronal jets can form at the periphery of active regions by reconnection between emerging, closed fields and pre-existing, semi-open active-region fields. \citet{archontis10} showed that recurrent jets are produced in this simulation if the flux emergence persists for a sufficiently long time.

\articlefigure[width=1.0\textwidth]{\figspath/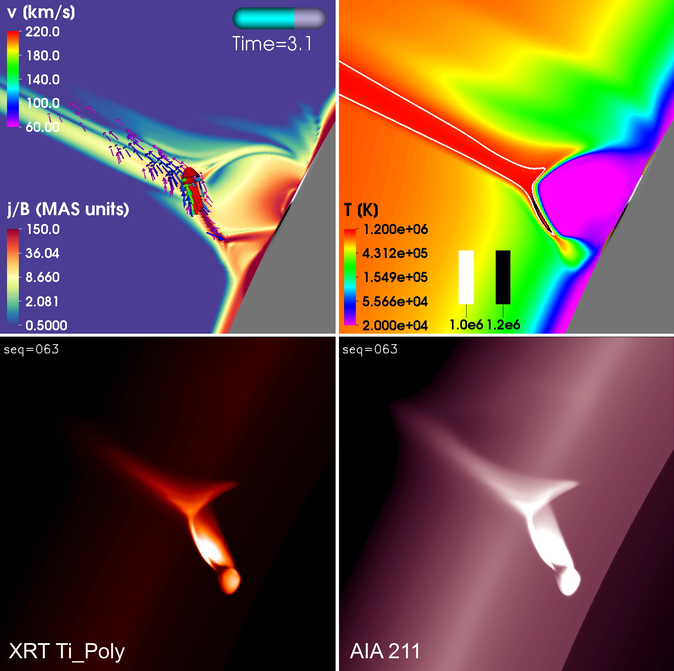}
{f:fig1}
{Thermodynamic MHD simulation of a standard jet.
\emph{Top left:} $j/B$ in normalized code units (outlining electric 
currents) and velocity vectors (arrows) in a vertical plane cutting 
through the flux-emergence region. 
\emph{Top right:} Plasma temperature in the same plane. 
\emph{Bottom left} and \emph{right:} Synthetic emission images showing 
the inverted-Y shape typically observed in coronal jets.}

\citet{pariat09} presented a different approach, in which a jet is produced without flux emergence. Their work was motivated by the conjecture that the build-up of current sheets on the slow time-scales of flux emergence cannot account for the impulsive nature of coronal jets, since the current sheets will dissipate before a sufficient amount of free magnetic energy accumulates. They started with a fan-spine magnetic configuration that was twisted below the fan by photospheric flows. Since reconnection is forbidden in such a configuration as long as the photospheric flux is preserved, it sets in only after the symmetry of the system is broken (by a kink-like instability), thereby resulting in an impulsive energy release. The reconnection between the twisted closed and untwisted open field lines launches a torsional Alfv\'en wave which compresses and accelerates the plasma, yielding an upwardly directed, helical jet that exhibits unwinding motions. The model can also account for recurrent jets if the twisting is continued after the launch of the first jet \citep{pariat10}. In a subsequent study \citet{rachmeler10} showed that no jet occurs if reconnection is inhibited, supporting the idea that reconnection is indeed essential for jet formation. 

\articlefigure[width=1.0\textwidth]{\figspath/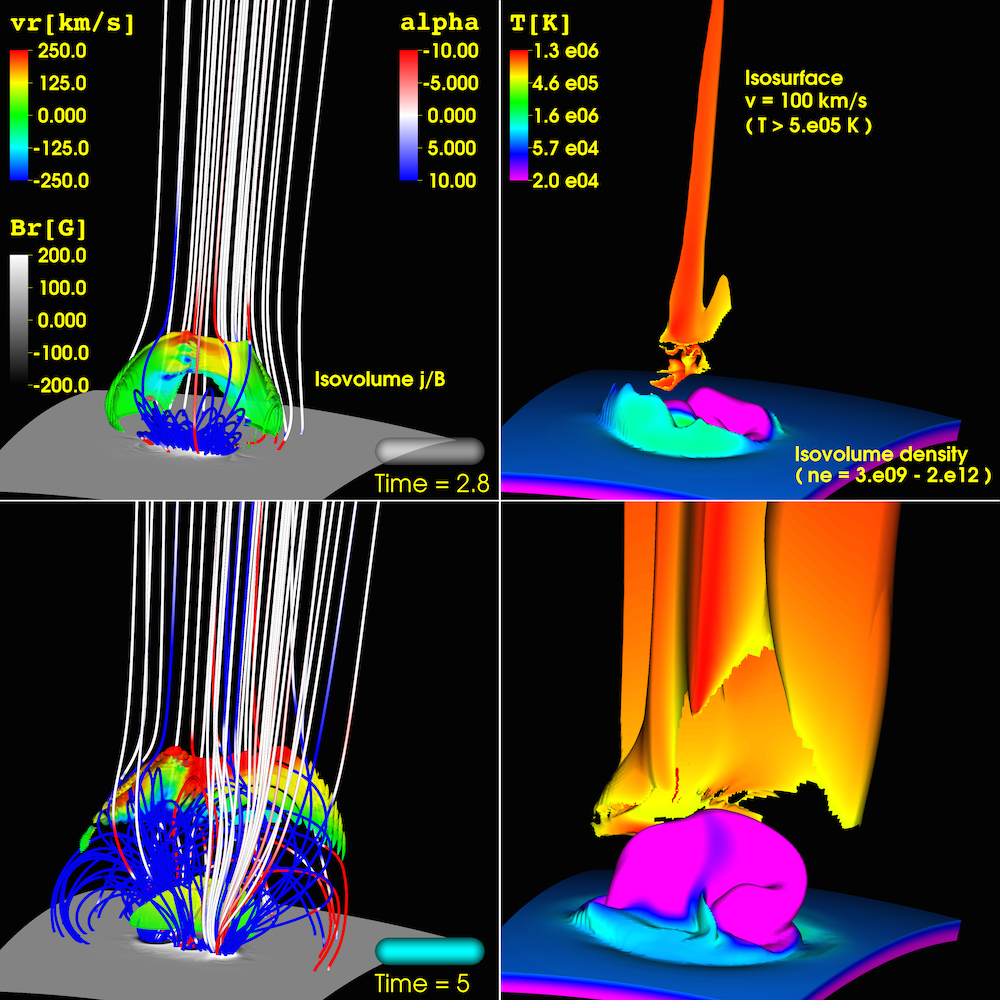}
{f:fig2}
{Successive occurrence of a standard jet {\em (top)} and a blowout jet 
{\em (bottom)} in a simulation where flux emergence was imposed for
a longer time than in the one shown in Figure\,\ref{f:fig1}. 
{\em Left:} Magnetic field lines colored by $\alpha={\bf j} 
\cdot {\bf B} / B^2$ in normalized units (outlining twist), radial magnetic 
field, $B_r$, at the bottom boundary, and iso-volumes showing current layers 
colored by radial velocity, $v_r$. {\em Right:} Iso-volumes of density and 
iso-surfaces of $v_r$ (drawn where $T>0.5$\,MK), colored by 
temperature.}

The latest development in the MHD modeling of coronal jets was provided by \cite{archontis13} and \cite{moreno-insertis13}. Both groups simulated the transition from a standard to a blowout jet by running their 3D simulations of flux emergence into an oblique coronal field in a larger domain and for a longer time. The simulations demonstrate how the eruption of a part of the emerging flux rope leads to the widening of the jet spire, the formation of a growing arcade of hot loops below the erupting flux rope, and the ejection of cool chromospheric material, as suggested by \citet[][]{moore10}. Very recently, \cite{fang14} performed a similar simulation and found that the inclusion of thermal conduction increases the mass flow into the corona and improves the agreement between simulated plasma emission and the emission morphology expected from the underlying magnetic field structure.  

These simulations were able to reproduce the morphology, velocities, and basic plasma properties of standard and blowout coronal jets in reasonably good agreement with the observations. However, despite considerable progress, there are various aspects that have not yet been fully addressed. For example, (i) the respective contributions of the physical mechanisms that seem to play a role in heating and accelerating the jet plasma (e.g., chromospheric evaporation, heat conduction, fast and slow-mode shocks, Lorentz and pressure gradient forces, torsional Alfv\'en waves) have not yet been clearly established. (ii) There exists no 3D simulation that properly models the energy transfer in the corona by taking into account thermal conduction, radiative losses, and background heating. Conduction and radiation were considered so far only in 1D and 2D \citep{shimojo01,miyagoshi03,miyagoshi04}, while \cite{yang.liping13b} also included background heating in a 2.5D simulation of chromospheric jets and \cite{fang14} included for the first time heat conduction in 3D simulations. (iii) All simulations published so far consider relatively small numerical domains in Cartesian geometry, modeling heights of only up to $\approx$\,70\,Mm above the solar surface. Hence, the possible mass and energy contribution of jets to the solar wind could be estimated only very roughly using simulations \cite[e.g.][]{lee15}. (iv) There exists no data-driven MHD simulation of an observed event. At present, only \cite{cheung15} aimed to model a real event, but they used the magneto-frictional approximation which has limited ability to describe the plasma dynamics. (v) Despite the fact that some, or even many, coronal jets appear to be triggered by flux cancellation rather than flux emergence \citep[e.g.,][]{chifor08,adams14,young14}, their formation and evolution has not yet been simulated in corresponding flux-convergence models such as, for example, the ones by \cite{javadi11} or \cite{yang.liping13b}. 

We have recently conducted simulations that will allow us to address items (i)--(iii). In the remainder of this article, we briefly describe our simulations and present preliminary results. A more detailed account will be given in forthcoming publications.

%
\section{Large-Scale Thermodynamic MHD Simulations of Coronal Jets}
%
Our simulations are performed using the MAS (Magnetohydrodynamic Algorithm outside a Sphere) code \citep[e.g.,][]{mikic94,linker99,lionello99,mikic99}. The code uses spherical coordinates and advances the standard viscous and resistive MHD equations. It incorporates radiative losses, thermal conduction parallel to the magnetic field, and an empirical coronal heating function. The latter properties are essential for a realistic modeling of the plasma densities and temperatures in the corona, and provide the possibility to produce synthetic EUV and soft X-ray images that can be compared directly to the observations \citep[see][]{lionello09}. The simulation domain covers the corona from 1--20 $R_\odot$, where $R_\odot$ is the solar radius. For simplicity we consider a purely radial coronal magnetic field, produced by a magnetic monopole located at Sun center. We include the solar wind and use a exponentially decaying coronal heating function \citep[see][]{lionello09}. 

We adopt the flux-emergence scenario to model jets in this background configuration. After a steady-state solution of the large-scale corona and solar wind is obtained, the emergence of a flux rope is modeled ``kinematically'', i.e., by successively imposing data at the lower boundary of the simulation domain, using the technique described in \cite{lionello13}. The data are extracted from a flux-emergence simulation similar to the runs ``ND'' and ``ND1'' described in \cite{leake13}. 

As the flux rope expands in the corona, a current layer is formed and reconnection across this layer triggers the development of a standard jet (Figure\,\ref{f:fig1}). The reconnection is driven by the continuous expansion of the flux rope and occurs in episodic bursts. We find that its onset does not launch a jet immediately, since the plasma density in and around the current layer is too large to be heated efficiently at this time \citep[see also][]{moreno-insertis13}. Only after mass evacuation by means of a ``plasmoid'' ejection takes place, does the subsequent plasma compression driven by the increase of thermal pressure raise the temperature to values above 1\,MK, and a jet is formed. Figure\,\ref{f:fig1} shows the resulting temperature distribution (note that ohmic heating, which is expected to lead to higher temperatures, was switched off for all runs shown in this article -- its influence will be studied separately). The synthetic emission images obtained at this time show the inverted-Y shape characteristic for standard jets. 

\articlefigure[width=1.0\textwidth]{\figspath/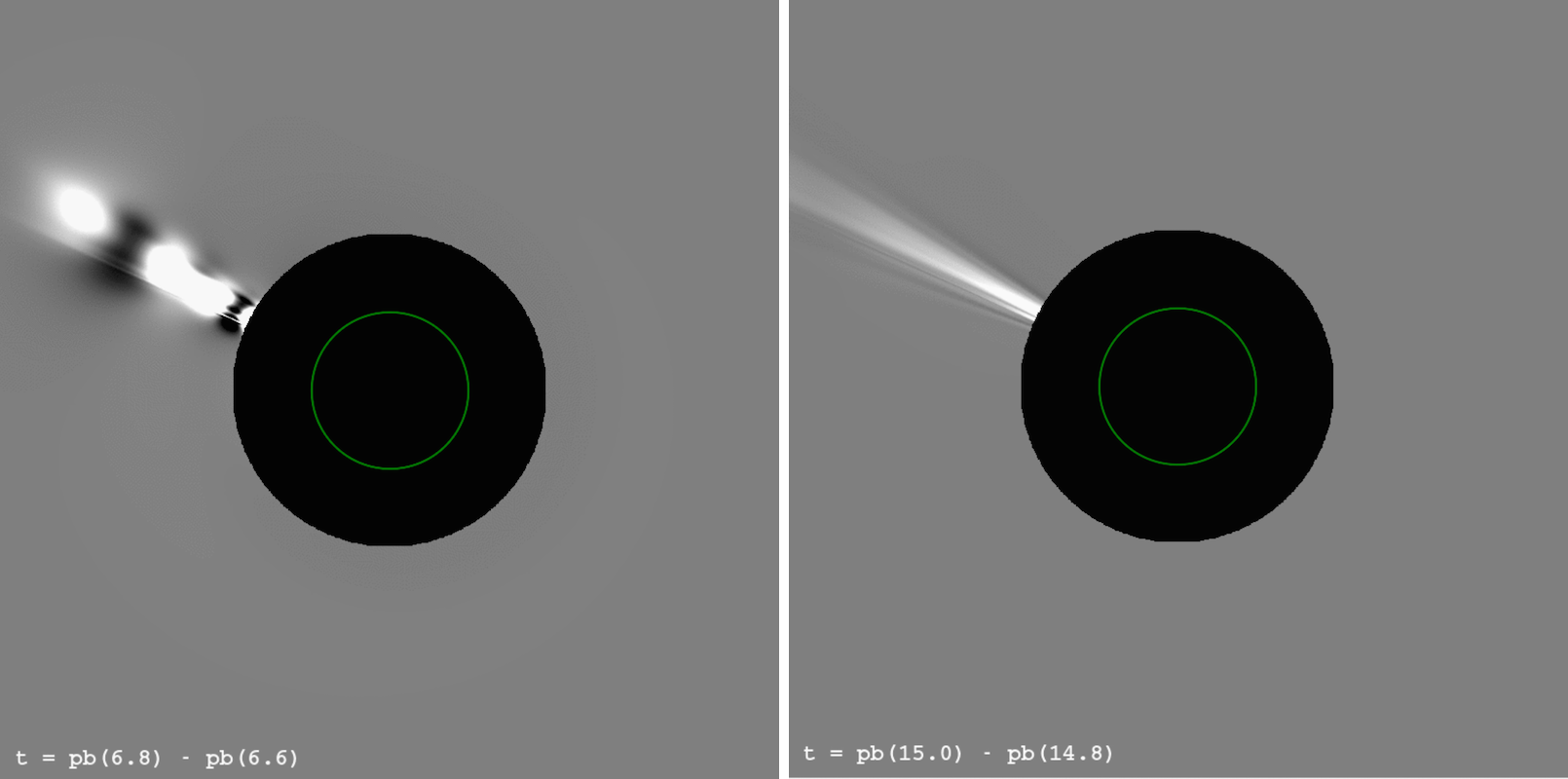}
{f:fig3}
{Synthetic white-light images of the corona (running-difference of polarization
brightness), 2.0 hours {\em (left)} and 5.3 hours {\em (right)} after the launch 
of a standard jet. 
The green circle marks the solar surface, the field-of-view of the artificial 
coronagraph is (2-5) $R_\odot$. The left panel shows ``plasma blobs'' originating 
from episodic reconnection outflows during the jet, the right panel shows a 
plume-like structure that develops after reconnection has ceased and the blobs 
have left the field-of-view.}

If flux emergence is imposed for a sufficiently long time, as for the simulation shown in Figure\,\ref{f:fig2}, the standard jet is followed by a significantly more energetic blowout jet, which occurs when the emerging flux rope becomes partially unstable and erupts. The figure illustrates that the region of fast plasma up-flow is much wider in the blowout jet, and that cold and dense plasma is ejected together with the erupting flux rope.

\articlefigure[width=1.0\textwidth]{\figspath/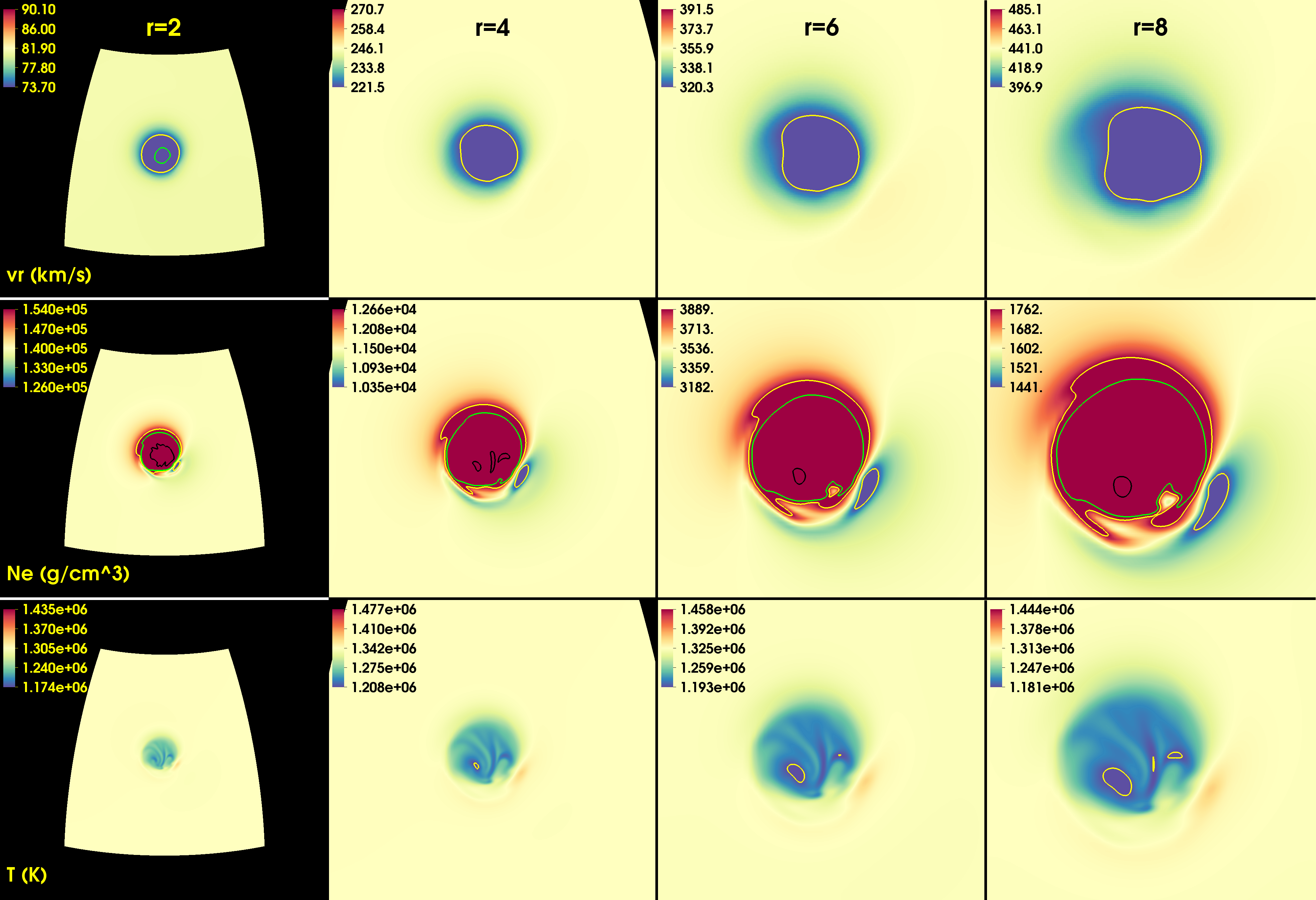}
{f:fig4}
{Radial velocity, density, and temperature at four different heights
in the corona for the same simulation as in Figure\,3, shown here 9.4 
hours after the launch of the jet. The view is onto the center of the 
emerging flux region. Color scales range from 90\,\% to 110\,\% of the 
respective initial values. Yellow, green, and black contour lines 
correspond to an increase/decrease of 10, 20, and 80\,\%, respectively.}   

Figures\,\ref{f:fig3} and \ref{f:fig4} depict a standard-jet simulation in the outer corona. This simulation is similar to the ones shown in Figures\,\ref{f:fig1} and \ref{f:fig2} but was ran for a longer time (for $\approx\,10$\,h), and the occurrence of a flux-rope eruption and blowout jet was suppressed by using a sufficiently large resistivity. In this run, the emergence is imposed for 2.2 hours and the jet is launched after 0.7 hours. The synthetic white-light image in the left panel of Figure\,\ref{f:fig3}, produced two hours after the launch of the jet, shows the presence of several ``blobs'', which are a manifestation of episodic reconnection outflows at larger heights. These structures are reminiscent of those first reported in connection with CMEs by \cite{sheeley99}. A few hours later, after the blobs have left the field-of-view of the artificial coronagraph, a plume-like structure of enhanced white-light emission (i.e., plasma density) has developed (right panel of Figure\,\ref{f:fig3}). 

Figure\,\ref{f:fig4} shows the radial velocity, electron number density, and plasma temperature at four different heights in the corona by the end of the simulation, 9.4 hours after the launch of the jet. It can be seen that the jet has evolved to heights $\gtrsim 8\,R_\odot$, i.e., it perturbs the solar wind. The lateral extension of the perturbation remains localized, presumably largely channeled by the radial magnetic field. For this standard-jet simulation we find a clear increase of the plasma densities with respect to the surroundings, i.e., the jet supplies mass to the solar wind. On the other hand, the plasma flows and temperatures are smaller than those in the ambient solar wind. The values of these quantities, however, may depend strongly on the respective reconnection outflow speeds and plasma temperatures produced during the jet. The latter, in turn, can be expected to depend on model parameters such as the strength of the emerging flux and whether or not ohmic heating is included. We will systematically investigate the influence of such model parameters, and the resulting contributions of both standard and blowout jets to the outer corona and solar wind in future publications.

%
\section{Summary}
%
Observations and numerical simulations have significantly improved our knowledge of coronal jets in the past two decades. Still, many aspects of coronal jets, such as the respective roles of the underlying physical mechanisms and the potential contribution of these events to coronal heating and the solar wind, are not yet well understood. A deeper insight can be provided in the near future by a ``next generation'' of jet simulations which employ thermodynamic MHD in large-scale coronal domains that include the modeling of the solar wind \citep[see also][]{karpen15}. This article provided a first impression of the capabilities of such simulations.

\acknowledgments
The plots in Figures\,1, 2, and 4 were created using the Open Source visualization tool {\em VisIt} {\small (\url{https://wci.llnl.gov/simulation/computer-codes/visit/})}. Computational resources were provided by the NSF-supported Texas Advanced Computing Center (TACC) in Austin, the NASA Advanced Supercomputing Division (NAS) at Ames Research Center, and the DoD High Performance Computing Program. TT, RL, VST, ZM, and JAL were supported by NASA's LWS and HSR programs. JEL and MGL were supported by NASA's HSR Program, and by a grant form the Chief of Naval Research. 


\end{document}